\documentclass[12pt]{iopart}

\usepackage {graphicx}
\begin{document}
\title{Reliability and efficiency of generalized rumor spreading model on complex social networks}
\author{Yaghoob Naimi$^{1}$ and  Mohammad Naimi$^{2}$}

\address{$^{1}$ Department of Physics, Lamerd Branch, Islamic Azad University, Lamerd, Iran.}
\address{$^{2}$ Department of Computer, Lamerd Branch, Islamic Azad University, Lamerd, Iran.}
\ead{y.naimi@iaulamerd.ac.ir and mohammad\_n60@iaulamerd.ac.ir}
\begin{abstract}
We introduce the generalized rumor spreading model and
investigate some properties of this model on
different complex social networks. Despite pervious rumor models that both the spreader-spreader ($SS$)
and the spreader-stifler ($SR$) interactions have the same rate $\alpha$, we define $\alpha^{(1)}$ and $\alpha^{(2)}$ for $SS$ and $SR$ interactions, respectively. The effect of variation of $\alpha^{(1)}$ and $\alpha^{(2)}$ on the final density of stiflers is investigated. Furthermore, the influence of the
topological structure of the network in rumor spreading is studied by analyzing the behavior of
several global parameters such as reliability and efficiency. Our results show that while networks with homogeneous connectivity patterns reach a higher reliability, scale-free topologies need a less time to reach a steady state with respect the rumor.
\end{abstract}
\noindent{PACS number(s): 89.75.Hc, 02.50.Ey, 64.60.aq}
\section{Introduction}
Network modeling is the recent interest of a wide interdisciplinary academic field which studies complex systems such as
social, biological and physical systems \cite{b.1, b.2}. By using a networked representation, it is possible to compare, in the same framework,
systems that are originally very different, so that the identification of some universal properties becomes much easier. Moreover, a network description of complex system allows to obtain related information by means of completely statistical coarse-grained analyses, without taking into account the detailed characterization of the system. So universality and simplicity are two fundamental principles that are interested in the study of the emergence of collective phenomena in systems with many interacting components.

From a general point of view, complex networks are connected graphs with, at most, a single edge between nodes where nodes stand for individuals and an edge corresponding to the interaction between individuals \cite{b.3, b.4}. The collective behavior of nodes is complex in the sense that it can not be directly predicted and characterized in terms of the behavior of each individual. The collective behavior is the responsible of interactions that occurs when pairs of components are connected with links. It is simple to find various systems in both nature and society that can be described in this manner. The most studied class of network modelling is the communication networks such as the Internet \cite{b.5} and the World Wide Web \cite {b.6}. A second class is related to social networks such as sexual contact networks \cite{b.7}, friendship networks \cite{b.8} and scientific collaboration networks \cite{b.9}. The last large class is concerned to biological networks such as metabolic networks \cite{b.10} and food webs \cite{b.11}.

Rumors have been a basic element of human interaction for as long as people have had questions about their social environment \cite{b.12}. Rumors are known for spreading between people quickly and easily since they are easy to tell, but hard to prove. Sometime, a rumor contains harmful information, so it is impossible to ignore, and can has damaging and perhaps even deadly consequences. We know it is probably bad for us, and we know it can hurt those around us, but we often find it hard to resist becoming active participants in the rumor spreading process. In business settings, it can greatly impact financial markets \cite{b.13, b.14}.

Despite its obvious negative connotations, a rumor has the capacity to satisfy certain fundamental personal and social needs and can shape the public opinion in a country \cite{b.12}. To a great extent, rumors help people make sense of what is going on around them. In this case, rumors spreading becomes a means by which people try to get the facts, to obtain enough information so that it reduces their psychological discomfort and relieves their fears.

A rumor can be interpreted as an infection of the mind. Daley and Kendall (DK) have introduced the original model of rumor spreading \cite{b.15, b.16}. In the DK model a closed and homogeneously mixed population can be classified into three distinct classes. These classes are called ignorants, spreaders and stiflers. The ignorants, those who have not heard the rumor yet, so they are susceptible to become infected by rumor.
The second class consists of the spreaders, those who have heard the rumor and are still interested to transmit it. Eventually, the stiflers,
those who have heard the rumor but have lost interest in the rumor and have ceased to transmit it. When the pairwise contacts between spreader and others occur in the society, the rumor is propagated through the population. If a spreader meets an ignorant, the last
one turns into a new spreader with probability $\lambda$; otherwise, the spreader meets another spreader or stifler, so they conclude that the rumor is known and do not spread the rumor anymore, therefore, turning into stiflers with probability $\alpha$. An important variant model of DK is the Maki-Thompson (MK) model \cite{b.17}. In the MK model, the rumor is spread by directed contacts of the spreaders with
others. Furthermore, in the contacts of the type spreader-spreader, only the initiating spreader becomes a stifler. Therefore, there is no double
transition to the stifler class. In the past, the DK and the MK models were used extensively to study rumor spreading \cite{b.18,b.19,b.20}.

In the above-mentioned models of rumor spreading, the authors have investigated the rumor spreading in the homogeneous networks that their degree distributions are very peaked around the average value, with bounded fluctuations \cite{b.21,b.22}. While, in the last years, a huge amount of experimental data yielded undoubtful evidences that real networks present a strong degree heterogeneity, expressed by a broad degree distribution \cite{b.23,b.24}. Recently, the model that we call the standard rumor model has
been studied in Ref. \cite{b.25} where authors studied a new model of rumor spreading on complex networks which, in comparison with previous models, provides a more realistic description of this process.
In standard rumor model unlike previous rumor models that stifling process is the only mechanism that results in cessation of rumor spreading, authors assumed two distinct mechanisms that cause cessation of a rumor, stifling and forgetting. In reality, cessation can occur also purely as a result of spreaders forgetting to tell the rumor, or their disinclination to spread the rumor anymore. They took forgetting mechanism into account by assuming that individuals may also cease spreading a rumor spontaneously (i.e., without any contact) with probability $\delta$. Furthermore, in the standard rumor spreading model, each node has an infectivity equal to its degree, and connectivity is uniform across all links. The generalization of the standard rumor model considered in Ref. \cite{b.25} has been studied in Ref. \cite{b.26} by introducing an infectivity function that determines the number of simultaneous contacts that a given node (individual) may establish with its connected neighbors and a connectivity strength function for the direct link between two connected nodes. These lead to a degree-biased propagation of rumors. To read more about social networks, one can refer to \cite{b.27,b.28,b.29,b.30}.

In the above-mentioned models of rumor spreading, it has been assumed that in both interactions of stifling process (i.e., spreader-spreader and spreader-stifler) the initiating spreader becomes a stifler with the same rate $\alpha$. In this paper we leave this assumption and assign a distinct rate for each interaction. More precisely, we introduce the generalized model in which the encounter of the spreader-spreader (spreader-stifler) leads to the stifler-spreader (stifler-stifler) with rate $\alpha^{(1)}$ ($\alpha^{(2)}$). We study in detail the dynamics of a generalized rumor model on some complex networks through analytic and numerical studies, and investigate
the impact of the interaction rules on the efficiency and reliability of the rumor process. The rest of the paper is organized as follows. In Section 2,
we introduce the standard model of rumor spreading and
shortly review epidemic dynamics of this model. In Section 3 we introduce the generalized rumor spreading model and analytically study the dynamics of this model on complex social networks in detail. The influence of the topological structure of the network in rumor spreading is studied by analyzing the behavior of
several global parameters such as reliability, efficiency in Section 4. Finally, our conclusions are presented in the last Section.

\section{Standard rumor spreading model}
\subsection{Definition of model}
The rumor model is defined as follows. Each of the
individuals (the nodes in the network) can be classified in three distinct
states with respect to the rumor as $I$, the ignorant or the individual has not
heard the rumor yet, $S$, the spreader or the individual is
aware of the rumor and is willing to transmit it, and $R$,  the stifler or the individual has heard the rumor
but has lost the interest in it, and does not transmit it anymore. Based on Maki and
Thompson model \cite {b.17}, the directed contact between spreaders and the rest of the population
is the main requirement for spreading the rumor. From mathematical point of view, these contacts only can occur along the links of an undirected graph $G(N,E)$, where $N$ and $E$ denote the nodes and the edges of the graph, respectively. The model that
we call the standard model has been studied in Ref. \cite{b.25}. By
following \cite{b.25}, the possible
processes that can occur between the spreaders and the rest of the
population are
\begin{itemize}
  \item spreading process: $SI\longrightarrow SS$ whenever a spreader meets an ignorant, the
ignorant becomes a spreader at a rate $\lambda$.
  \item stifling processes:\begin{itemize}
                             \item $SS\longrightarrow RS$ when a spreader contacts another spreader,
the initiating spreader becomes a stifler at a rate $\alpha$.
                             \item $SR\longrightarrow RR$ when a spreader encounters a stifler,
the spreader becomes a stifler at a rate $\alpha$.
                           \end{itemize}
  \item forgetting process:$S\longrightarrow R$ there is a rate $\delta$ for a spreader to forget
spreading a rumor spontaneously (i.e., without any contact).
\end{itemize}

\subsection{Dynamics of standard rumor model}
The individuals in social complex networks not only be in three different states
but also belong to different connectivity (degree) classes $k$, therefore we denote $I_{k}(t)$, $S_{k}(t)$ and $R_{k}(t)$ for densities of the ignorant, spreader, and stifler nodes (individuals) with connectivity $k$ at time $t$, respectively. These quantities satisfy the normalization condition $I_{k}(t)+S{k}(t)+R_{k}(t) = 1$ for all $k$ classes.
We shortly review some classical results of standard model, where Nekovee et
al. described a formulation of this model on networks in terms of
interacting Markov chains, and used this framework to derive, from
first-principles, mean-field equations for the dynamics of rumor
spreading on complex networks with arbitrary degree correlations
as follows: \begin{equation}\label{1} \frac{dI_{k}(t)}{dt}=-k\lambda
I_{k}(t)\sum_{l} S_{l}(t)P(l|k)
 \end{equation}
\begin{equation}\label{2} \frac{dS_{k}(t)}{dt}=k\lambda I_{k}(t)\sum_{l}
S_{l}(t)P(l|k)- k \alpha S_{k}(t)\sum_{l}
(S_{l}(t)+R_{l}(t))P(l|k)-\delta S_{k}(t)\end{equation}
\begin{equation}\label{3}\frac{dR_{k}(t)}{dt}= k \alpha S_{k}(t)\sum_{l}
(S_{l}(t)+R_{l}(t))P(l|k)+\delta S_{k}(t) \end{equation}
where the conditional probability $P(l|k)$ means that a
randomly chosen link emanating from a node of degree $k$ leads to a node of degree $l$.
Moreover, we suppose that the degrees of nodes in the whole
network are uncorrelated, i.e., $P(l|k)=lp(l)/\langle k
\rangle$ where $p(k)$ is the degree distribution and $\langle k
\rangle$ is the average degree. They have used approximate analytical and exacted
numerical solutions of these equations to examine both the steady-state and the time-dependent
behavior of the model on several models of social networks such as homogeneous networks, random graphs and
uncorrelated scale-free (SF) networks. They have found that, as a function of the rumor spreading rate, their model shows a new critical behavior on
networks with bounded degree fluctuations, such as random graphs, and that this behavior is absent in SF
networks with unbounded fluctuations in node degree distribution. Furthermore, the initial spreading rate at
which a rumor spreads is much higher in SF networks as compared to random graphs.

In standard model the authors have mainly focused on critical threshold in several models of social networks but in the following section we introduce generalized rumor model and we concentrate on the final fraction of the population
that heard the rumor, $R$, when the spreading rate is fixed, $\lambda=1$, and we vary the value of other parameters of our model.
\section{Generalized rumor model}
In the standard rumor model \cite{b.25}, authors have assumed that
both stifling processes, the $SS\longrightarrow RS$ and the $SR\longrightarrow RR$, have the same rate
$\alpha$. But in this paper, we leave this assumption and define
$\alpha^{(1)}$ and $\alpha^{(2)}$ for $SS\longrightarrow RS$ and $SR\longrightarrow RR$ interactions,
respectively. We will show that this separation of rates leads to
notable results. The other interactions and their rates of our
model are the same as the standard model. Now, the mean-field rate equations can be rewritten as
\begin{equation}\label{12}
\frac{dI_{k}(t)}{dt}=-\frac{k\lambda}{\langle k\rangle}
I_{k}(t)\sum_{l} lP(l)S_{l}(t)
\end{equation}
\begin{eqnarray}
\frac{dS_{k}(t)}{dt}&=&\frac{k\lambda}{\langle k\rangle} I_{k}(t)\sum_{l}
lP(l)S_{l}(t)- \frac{k \alpha^{(1)} }{\langle k\rangle}S_{k}(t)\sum_{l}
lP(l)S_{l}(t)\cr&-&\frac{k \alpha^{(2)}}{\langle k\rangle} S_{k}(t)\sum_{l}
lP(l)R_{l}(t)-\delta S_{k}(t)
\end{eqnarray}
\begin{eqnarray}
\frac{dR_{k}(t)}{dt}&=&\frac{ k \alpha^{(1)}}{\langle k\rangle} S_{k}(t)\sum_{l}
lP(l)S_{l}(t)\cr&+&\frac{k \alpha^{(2)}}{\langle k\rangle} S_{k}(t)\sum_{l}
lP(l)R_{l}(t)+\delta S_{k}(t). \end{eqnarray}
Eq. (4) can be
integrated exactly to yield: \begin{equation}\label{9}
I_{k}(t)=I_{k}(0)e^{-\frac{\lambda k}{\langle k
\rangle}\phi(t)},\end{equation} where $I_{k}(0)$ is the initial density
of ignorant nodes with connectivity $k$, and we have used the
auxiliary function \begin{equation}\label{10}
\phi(t)=\sum_{k}k p(k)\int_{0}^{t}S_{k}(t')dt'\equiv
\int_{0}^{t}\langle k S_{k}(t')\rangle dt'.\end{equation}
In order to get a closed relation for finding the final fraction of the
population that heard the rumor, $R$, it is more useful to focus
on the time evolution of $\phi(t)$. Assuming an homogeneous
initial distribution of ignorant, i.e., $I_{k}(0)=I_{0}$ (without
lose of generality, we can put $I_{0}\approx 1$). The spreading process starts with one element becoming
informed of a rumour and terminates when no spreaders are left in the
population, i.e., $S_{k}(\infty)=0$ thus according to normalization condition, at the end of the epidemic we have
\begin{equation}\label{12}
R_{k}(\infty)=1-I_{k}(\infty)=1-e^{-\frac{\lambda k}{\langle k
\rangle}\phi(\infty)}
\end{equation}
After rather lengthy calculations, similar to what has been done in Refs. \cite{b.25,b.26}, one can find the expansion of $\phi(\infty)$ as
\begin{equation}\label{13}
  \phi(\infty)=\frac{\lambda\frac{\langle
k^{2}\rangle}{\langle
k\rangle}-\delta}{\lambda^{2}\frac{\langle
k^{3}\rangle}{\langle
k\rangle^{2}}\left[\frac{1}{2}+\alpha^{(1)}\delta^{2}\frac{\langle
k^{2}\rangle}{\langle
k\rangle}I^{2}+\alpha^{(2)}\delta I\frac{\langle
k^{2}\rangle}{\langle
k\rangle}(1-\delta I)\right]}
\end{equation}
where $I$ is a finite and positive integral that has the form $I=\frac{1}{\phi(\infty)}\int_{0}^{t}e^{\delta(t-t')}\phi(t')dt'$.
At the end of epidemic, the final fraction of the population that
heard the rumor, $R$, is given by \begin{equation}\label{19}
R=\sum_{k}p(k)(1-e^{-\frac{\lambda k}{\langle k
\rangle}\phi(\infty)}).\end{equation} Regardless of the network topology and configuration, for
any form of p(k), above relation can be simplified by expanding the
exponential for the first order in $\phi(\infty)$, one obtains \begin{equation}\label{20}
R\simeq \lambda\phi(\infty).\end{equation}
\section{Results and Discussions}
One of the most important practical aspects of any rumor
mongering process is whether or not it reaches a high
number of individuals that heard the rumor. This value is simply given
by the density of stiflers, $R$, at the end of the epidemic and is called "reliability" of
the rumor process. For obvious practical purposes,
any algorithm or process that emulates an effective spreading of a rumor will try to find the conditions that under these the reliability reaches as much as possible value. Another important quantity is the efficiency of the process which is the ratio between the reliability and the
load imposed to the network. Load means number of messages on average each node sending to its neighbors in order to propagate the rumor. For these purposes, one does not only want to have high reliability levels, but also the lowest possible cost in terms of network load. This is important in order to reduce the amount of processing power used by nodes participating in the spreading process.

In order to characterize this trade-off between reliability and cost, we use time as a
practical measure of efficiency. Similar to Ref. \cite{b.31}, we call a rumor process
less efficient than another if it needs more time to reach
the same level of reliability.

To illustrate the effect of separation of stifling process rate
$\alpha$ (for both $SS$ and $SR$ interactions) into $\alpha^{(1)}$
and $\alpha^{(2)}$ for $SS$ and $SR$, respectively, we consider a
standard scale-free (SF) and Erd\H{o}s-R\'{e}nyi (ER) network. The SF network has generated according
to $p(k)\sim k^{-3}$, the number of nodes is $N=4000$ and the
average degree is $\langle k \rangle=8$. The ER network is a homogenous network that has the size $N=4000$ with $\langle k \rangle=8$.
Throughout the rest of the paper we set $\lambda=1$ without loss of generality and vary
the value of $\alpha^{(1)}$ and $\alpha^{(2)}$. Fig. 1 shows the time evolution
of the density of spreaders for different values of the stifling process rates,
$\alpha^{(1)}$ and $\alpha^{(2)}$, when the forgetting process rate is $\delta=0.5$.

We define two different models according to variation of $\alpha^{(1)}$ and $\alpha^{(2)}$ as following
\begin{itemize}
  \item Model 1: $\alpha^{(1)}$ varies with condition \{$\alpha^{(1)}\leq\alpha^{(2)}$ and $\alpha^{(2)}$=1\},
  \item Model 2: $\alpha^{(2)}$ varies with condition \{$\alpha^{(2)}\leq\alpha^{(1)}$ and $\alpha^{(1)}$=1\}.
\end{itemize}

We have performed large scale numerical simulations by applying two stated conditions on SF and ER
networks. Figs .1 (Fig. 2) corresponding to model 1 (model 2) illustrates, as expected, that the number of individuals who spread the rumor increases as the stifling process rate $\alpha^{(1)}$ ($\alpha^{(2)}$) decreases. In the cases in which the $\alpha^{(1)}+\alpha^{(2)}$ is fixed, i.e., $(\alpha^{(1)},\alpha^{(2)})=(\alpha^{(2)},\alpha^{(1)})$, the maximum value of spreaders in Fig. 1 (a) (Fig. 2 (a)), the case in which $\alpha^{(1)}<\alpha^{(2)}$, is greater than the corresponding values in Fig.1 (b) (Fig. 2 (b)), the case in which $\alpha^{(2)}<\alpha^{(1)}$, although the lifetime of spreaders in latter is greater. On the other worlds, when the $\alpha^{(1)}+\alpha^{(2)}$ is fixed and $\alpha^{(1)}<\alpha^{(2)} $, more individuals participate in spreading the rumor. On the other hand, in model 1 the time it takes for $S(t)$ to reach its final value, i.e., no spreaders are left in the
population, very slightly varies with different amount of $\alpha^{(1)}$, but clear differences arise between the time that spreaders die out for different amount of $\alpha^{(2)}$ in model 2.

Fig. 3 (a) (Fig. 3 (b)) shows the final densities of stiflers for SF (ER) network. It obvious that in model 2 on both networks, the lower $\alpha^{(2)}$ leads to higher reliability (blue-solid curves). On the other hand, the time it takes for $R(t)$ to reach its asymptotic
value slightly increases when $\alpha^{(2)}$ decreases, the clear differences arise for the two lowest values of $\alpha^{(2)}$. Imposing the condition of model 1 on both networks, the red-dashed curves, illustrates that the lower $\alpha^{(1)}$ leads to higher reliability but unlike the previous case, the time it takes for $R(t)$ to reach its asymptotic value slightly decreases when $\alpha^{(1)}$ decreases as the inset figures show.

Generally, from Fig. 3, after the comparison of the cases in which $(\alpha^{(1)},\alpha^{(2)})=(\alpha^{(2)},\alpha^{(1)})$, we can conclude that the model 1 (blue-solid curves) leads to more reliable rumor spreading model but under condition of model 2 (red-dashed curves) the society reaches the steady state with respect the rumor in less time.

To make the comparison between the number of stiflers in the SF and ER networks at the end of epidemic, we have plotted the Fig. 4. As shows this figure, in the ER network the number of stiflers at the end of the process is definitely higher than SF network, so the SF network appears less reliable. It results that ER networks allow a larger reliability $R$ to this epidemic process. This is not straightforward and one may think that the existence of hubs in SF networks helps propagate the rumor. However, a closer look at the spreading dynamics reveals us that the presence of hubs introduces conflicting effects in the dynamics. While hubs may in principle contact with a larger number of individuals, spreader-spreader and spreader-stifler interactions get favored on the long run. More precisely, it is very likely that a hub in the spreader state turns into a stifler before contacts all its ignorant neighbors. Once a few hubs are turned into stiflers many of the neighboring individuals could be isolated and never get the rumor. In this sense, homogeneous networks allow for a more capillary propagation of the rumor, since all individuals contribute almost equally to the rumor spreading. On the other hand, the SF network reaches the steady state with respect the rumor
in less time than the homogeneous network (at the same condition). In this sense, SF network has a better efficiency.
\begin{figure}
\centering
\mbox{\includegraphics[height=2.5in,width=3in]{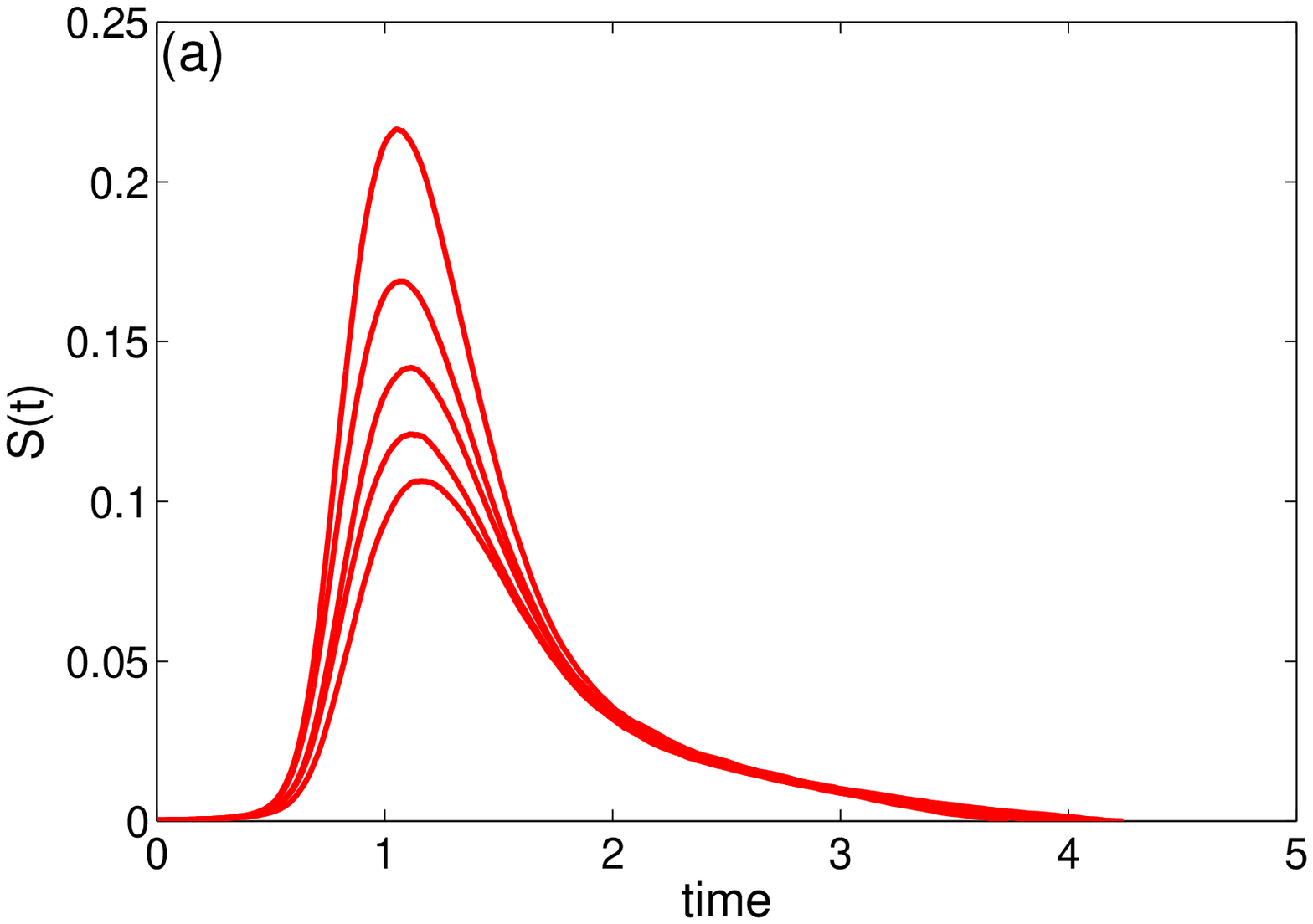}\quad
\includegraphics[height=2.5in,width=3in]{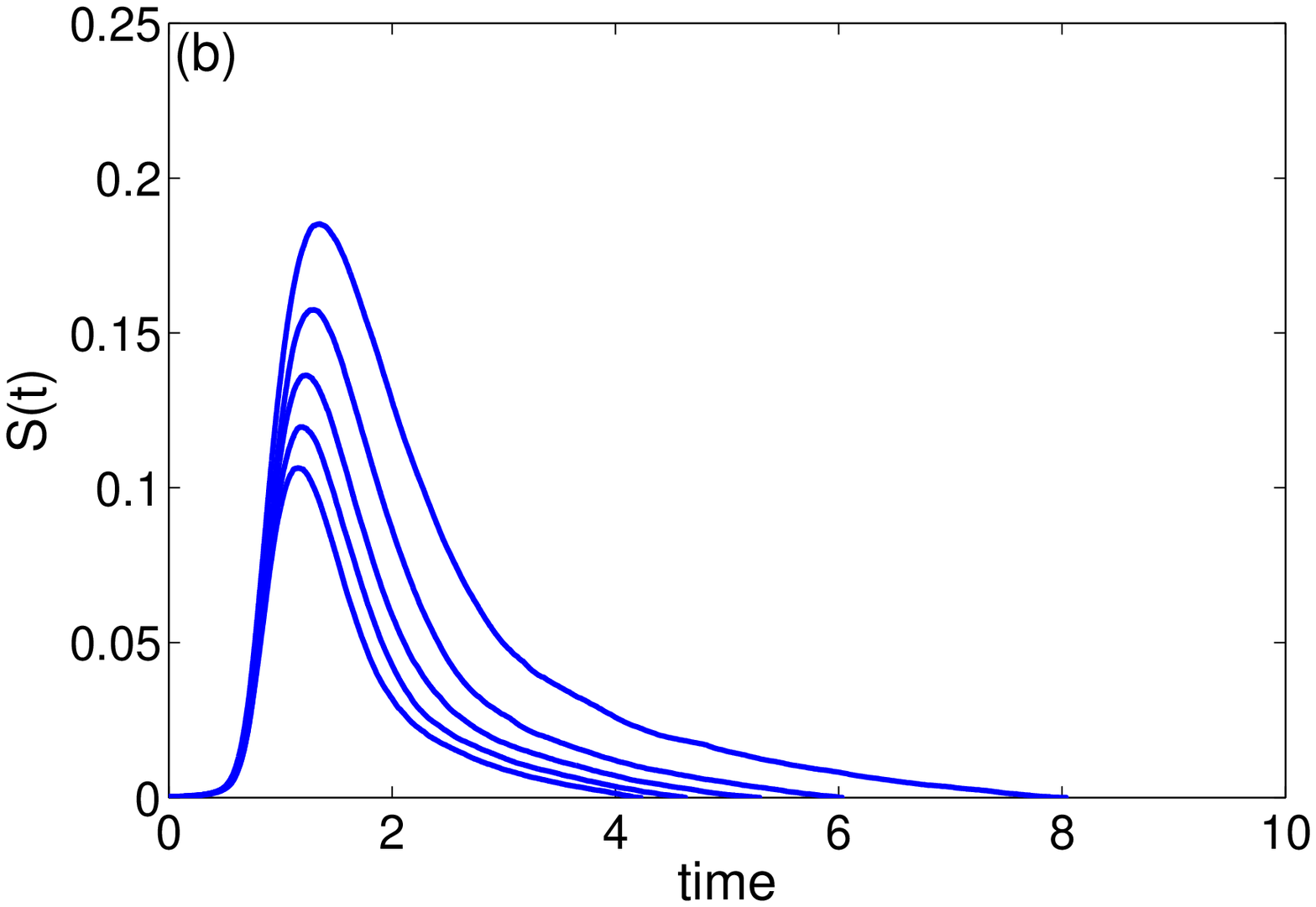} }
\caption{Time evolution of the density of spreaders in SF networks of size
$N=4000$ with $\langle k \rangle=8$ for (a) model 1 (b) model 2. In model 1 (model 2) from below $\alpha^{(1)}$ ($\alpha^{(2)}$) goes from 1.0 to 0.2 at fixed increments of 0.2.} \label{fig1}
\end{figure}
\begin{figure}
\centering
\mbox{\includegraphics[height=2.5in,width=3in]{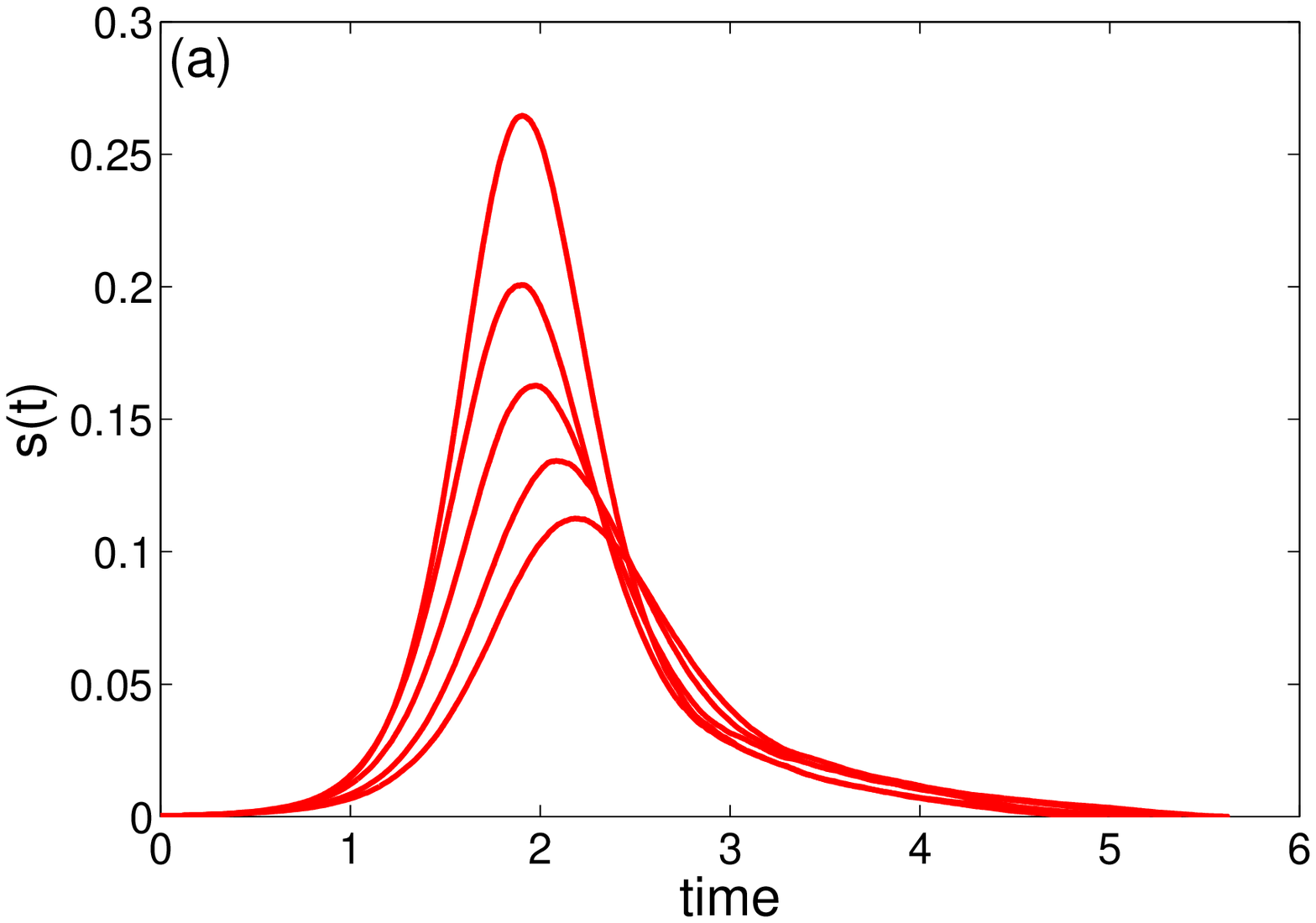}\quad
\includegraphics[height=2.5in,width=3in]{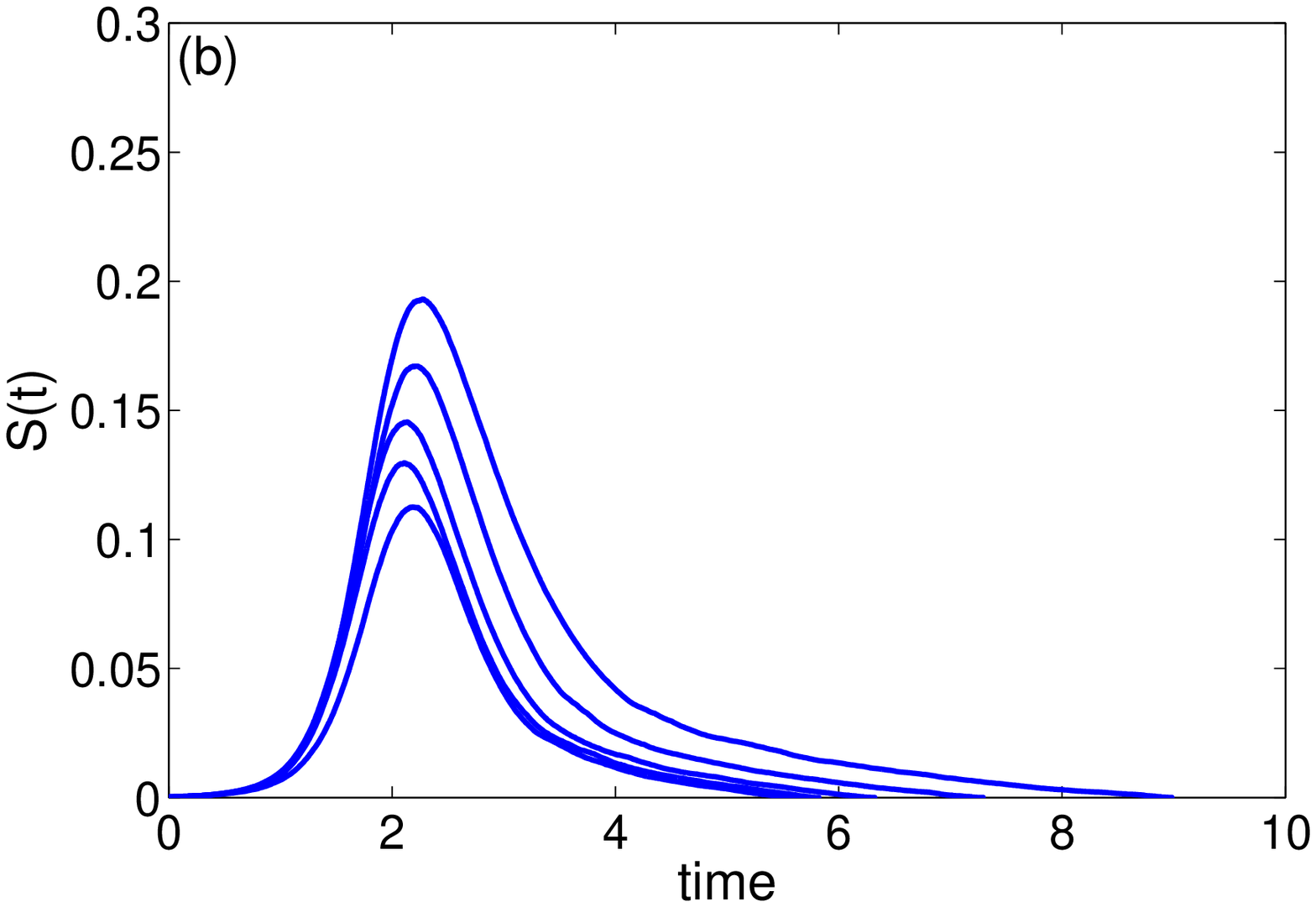} }
\caption{Time evolution of the density of spreaders in ER networks of size
$N=4000$ with $\langle k \rangle=8$ for (a) model 1 (b) model 2. In model 1 (model 2) from below $\alpha^{(1)}$ ($\alpha^{(2)}$) goes from 1.0 to 0.2 at fixed increments of 0.2.} \label{fig1}
\end{figure}
\begin{figure}
\centering
\mbox{\includegraphics[height=2.5in,width=3in]{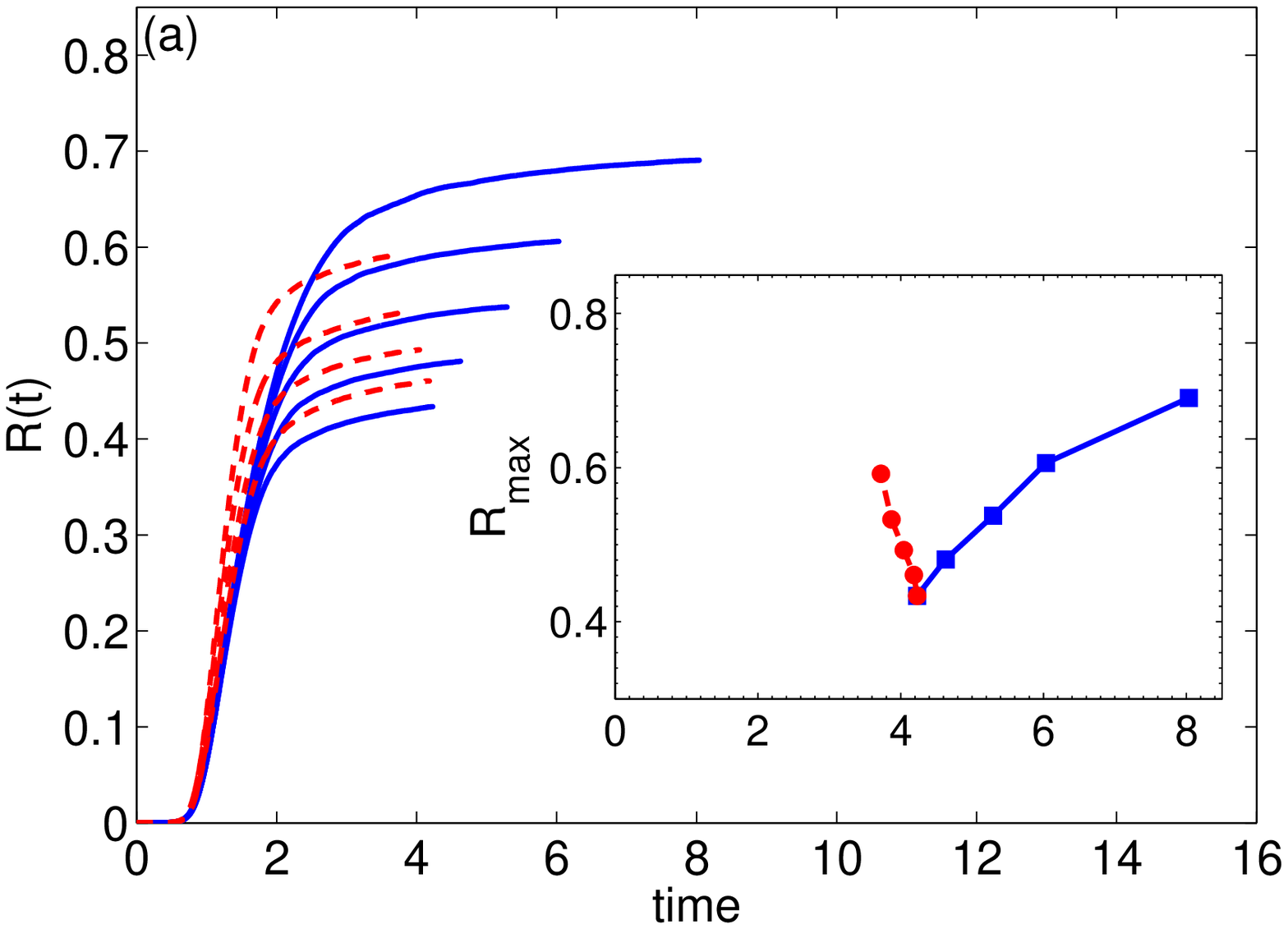}\quad
\includegraphics[height=2.5in,width=3in]{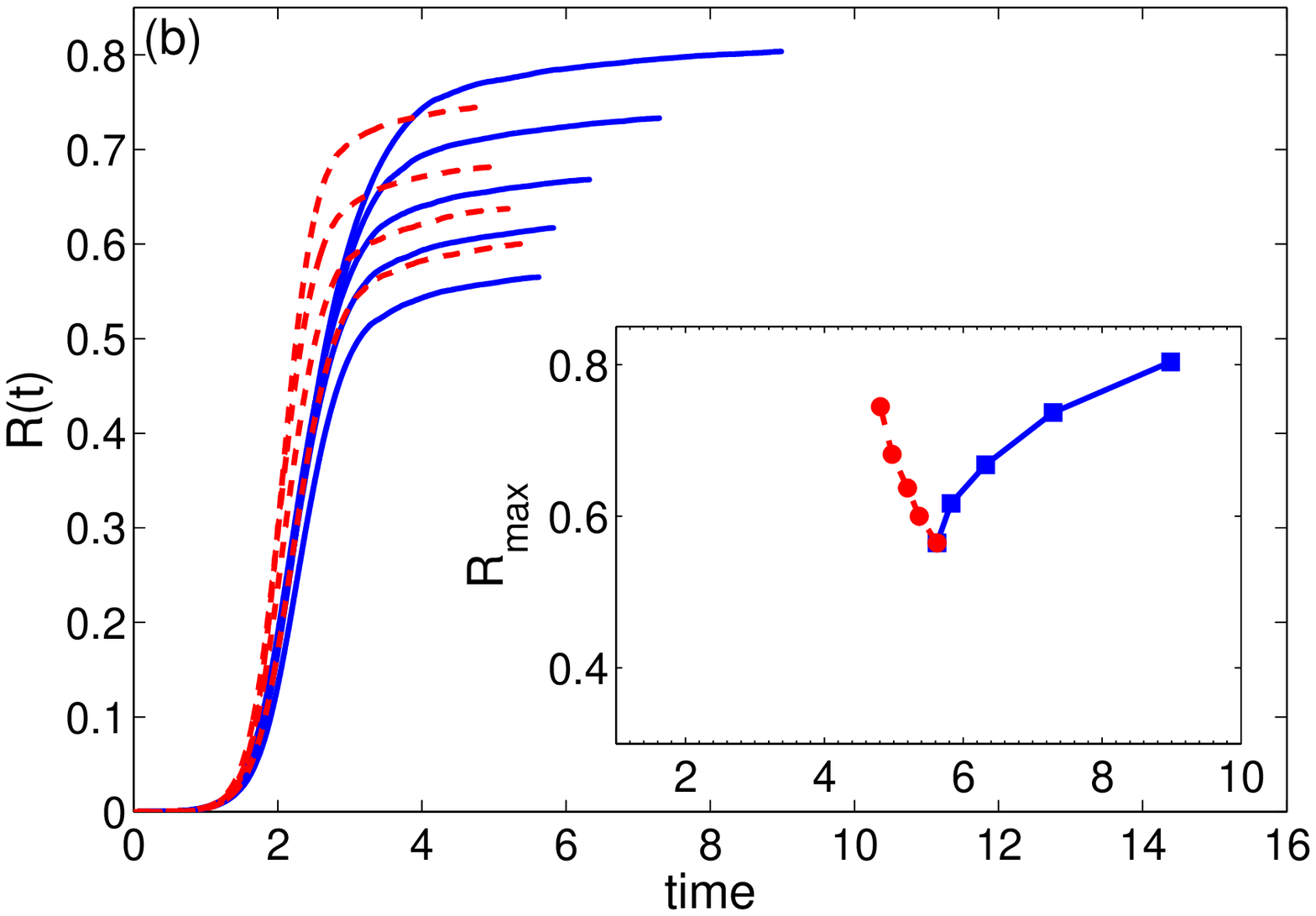} }
\caption{Time evolution of the density of stiflers for (a) SF network (b) ER network with the same size $N=4000$. The red-dashed and blue-solid curves are the stifler density corresponding to model 1 and model 2, respectively. The lowest curve in (a) and (b) is a common density for both models when $\alpha^{(1)}=\alpha^{(2)}=1$. From below, the variables go from 1.0 to 0.2 at fixed increments of 0.2. The insets show maximum value of stiflers (the endpoints of curves) at the end of the epidemic.} \label{fig2}
\end{figure}
\begin{figure}
\centering
\includegraphics[width=4in]{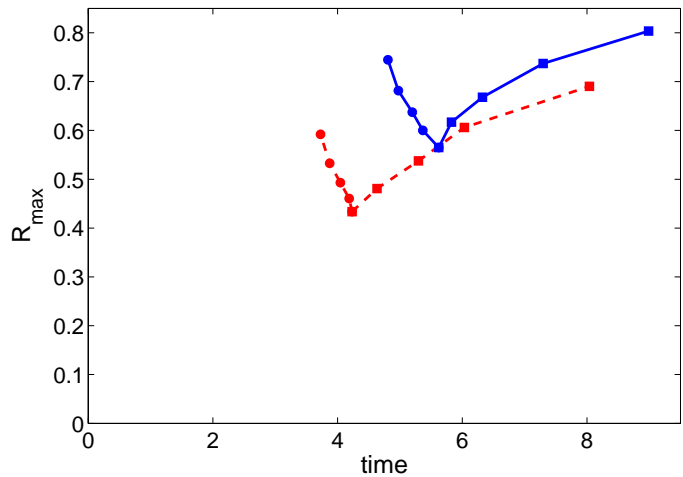}
\caption{Maximum value of stiflers at the end of epidemic for SF network (red-dashed) and ER network (blue-solid). The circles and squares show the maximum value of stifler with respect the time for model 1 and model 2, respectively. From below, the variables go from 1.0 to 0.2 at fixed increments of 0.2. } \label{fig2}
\end{figure}
\section{Conclusion}
In this paper we introduced a generalized model of rumor spreading on complex social networks. Unlike previous rumor models, our model incorporates two distinct rates for stifling processes. We have defined $\alpha^{(1)}$ and $\alpha^{(2)}$ for $SS\longrightarrow RS$ and $SR\longrightarrow RR$ interactions, respectively. Our simulations showed that in the condition $(\alpha^{(1)},\alpha^{(2)})=(\alpha^{(2)},\alpha^{(1)})$, when $\alpha^{(1)}$  is smaller than $\alpha^{(2)}$, the society reaches the steady state with respect the rumor in less time. On the other hand, when $\alpha^{(2)}<\alpha^{(1)}$ , the higher level of reliability is obtained. This result is valid for both homogeneous and heterogeneous (scale-free) networks.

By analyzing the behavior of several global parameters such as reliability and efficiency, we studied the influence of the topological structure of the network in
rumor spreading. Our results showed that while networks with homogeneous connectivity patterns reach a higher reliability, scale-free topologies need a less time to reach a steady state with respect the rumor.

\end{document}